\documentclass[sigconf, preprint]{acmart}
\usepackage{amsmath}
\usepackage{amsfonts}

\usepackage{amssymb}
\usepackage{amsthm}
\usepackage{booktabs}

\usepackage{graphicx}
\usepackage{booktabs}
\usepackage{bm}
\usepackage{hyperref}
\usepackage{xspace}
\usepackage{makecell}
\usepackage{multirow}
\usepackage{balance}
\usepackage{subfigure}
\usepackage{enumitem}
\usepackage{url}
\usepackage[flushleft]{threeparttable}
\usepackage{longtable}
\usepackage{supertabular}
\usepackage{microtype}
\usepackage[normalem]{ulem}

\RequirePackage{xspace}
\usepackage{nicefrac}       
\usepackage{microtype}      
\usepackage{lipsum}
\usepackage{natbib}

\usepackage{flushend}
\usepackage{enumerate}
\usepackage{subfigure}
\usepackage[T1]{fontenc}
\usepackage{framed}
\usepackage{stmaryrd}
\usepackage{xcolor}
\usepackage{enumerate}
\usepackage{enumitem}
\usepackage{graphicx}
\usepackage{hyperref}
\usepackage{wrapfig}
\usepackage{epstopdf}
\usepackage{url}
\usepackage[T1]{fontenc}
\usepackage{listings}
\usepackage{flushend}
\usepackage[ruled,vlined]{algorithm2e}
\usepackage{cleveref}
\usepackage{xargs}
\usepackage[colorinlistoftodos,prependcaption,textsize=tiny]{todonotes}
\usepackage[group-separator={,}, group-minimum-digits=4]{siunitx}

\newcommandx{\yli}[2][1=]{\todo[linecolor=blue,backgroundcolor=blue!25,bordercolor=blue,#1]{#2}}
\newcommandx{\improvement}[2][1=]{\todo[linecolor=yellow,backgroundcolor=yellow!25,bordercolor=yellow,#1]{#2}}
\newcommandx{\thiswillnotshow}[2][1=]{\todo[disable,#1]{#2}}

\newcommand{\blue}[1]{\textcolor{blue}{\small #1}}

\crefname{section}{§}{§§}
\Crefname{section}{§}{§§}

\AtBeginDocument{%
  }

\renewcommand\footnotetextcopyrightpermission[1]{}

\setcopyright{acmcopyright}
\copyrightyear{2018}
\acmYear{2018}
\acmDOI{XXXXXXX.XXXXXXX}

\acmConference[CIKM '23]{32nd ACM International Conference on Information and Knowledge Management}{August 6-10, 2023}{Long Beach, CA}

\begin{document}

\title{G-STO: Sequential Main Shopping Intention Detection via Graph-Regularized Stochastic Transformer}

\author{Yuchen Zhuang}
\authornote{Work done during the author's internship at Amazon.}
\email{yczhuang@gatech.edu}
\affiliation{%
  \institution{Georgia Institute of Technology}
  \city{Atlanta}
  \state{Georgia}
  \country{USA}
}

\author{Xin Shen}
\email{xinshen@amazon.com}
\affiliation{%
  \institution{Amazon}
  \city{Seattle}
  \state{Washington}
  \country{USA}
}

\author{Yan Zhao}
\email{yzhaoai@amazon.com}
\affiliation{%
  \institution{Amazon}
  \city{Seattle}
  \state{Washington}
  \country{USA}
}

\author{Chaosheng Dong}
\email{chaosd@amazon.com}
\affiliation{%
  \institution{Amazon}
  \city{Seattle}
  \state{Washington}
  \country{USA}
}

\author{Ming Wang}
\email{mingww@amazon.com}
\affiliation{%
  \institution{Amazon}
  \city{New York}
  \state{New York}
  \country{USA}
}

\author{Jin Li}
\email{jincli@amazon.com}
\affiliation{%
  \institution{Amazon}
  \city{Seattle}
  \state{Washington}
  \country{USA}
}

\author{Chao Zhang}
\email{chaozhang@gatech.edu}
\affiliation{%
  \institution{Georgia Institute of Technology}
  \city{Atlanta}
  \state{Georgia}
  \country{USA}
}

\renewcommand{\shortauthors}{Zhuang, et al.}
\newcommand{\ours}{\textsc{G-STO}\xspace}
\newcommand{\longours}{Graph-Regularized Stochastic Model\xspace}
\newcommand{\etal}{\emph{et~al.}\xspace} 
\newcommand{\etc}{\emph{etc.}\xspace} 
\newcommand{\ie}{\emph{i.e.}\xspace} 
\newcommand{\eg}{\emph{e.g.}\xspace} 
\begin{abstract}
Sequential recommendation requires understanding the dynamic patterns of users' behaviors, contexts, and preferences from their historical interactions.
Most existing works focus on modeling user-item interactions only from the item level, ignoring that they are driven by latent shopping intentions (\eg, ballpoint pens, miniatures, \etc).
Detecting these latent shopping intentions is crucial for e-commerce platforms such as Amazon to improve their customers' shopping experience.
Despite its significance, the area of main shopping intention detection remains under-investigated in the academic literature.
To fill this gap, we propose a graph-regularized stochastic Transformer method, \ours.
By considering intentions as sets of products and user preferences as compositions of intentions, we model both of them as stochastic Gaussian embeddings in the latent representation space.
Instead of training the stochastic representations from scratch, we develop a global intention relational graph as prior knowledge for regularization, allowing relevant shopping intentions to be distributionally close.
Finally, we feed the newly regularized stochastic embeddings into Transformer-based models to encode sequential information from the intention transitions.
We evaluate our main shopping intention identification model on three different real-world datasets, where \ours achieves significantly superior performances to the baselines by $18.08\%$ in Hit@1,  $7.01\%$ in Hit@10, and $6.11\%$ in NDCG@10 on average. 
\end{abstract}

\maketitle

\section{Introduction}

\begin{figure}[t]
\centering
\includegraphics[width=1\linewidth]{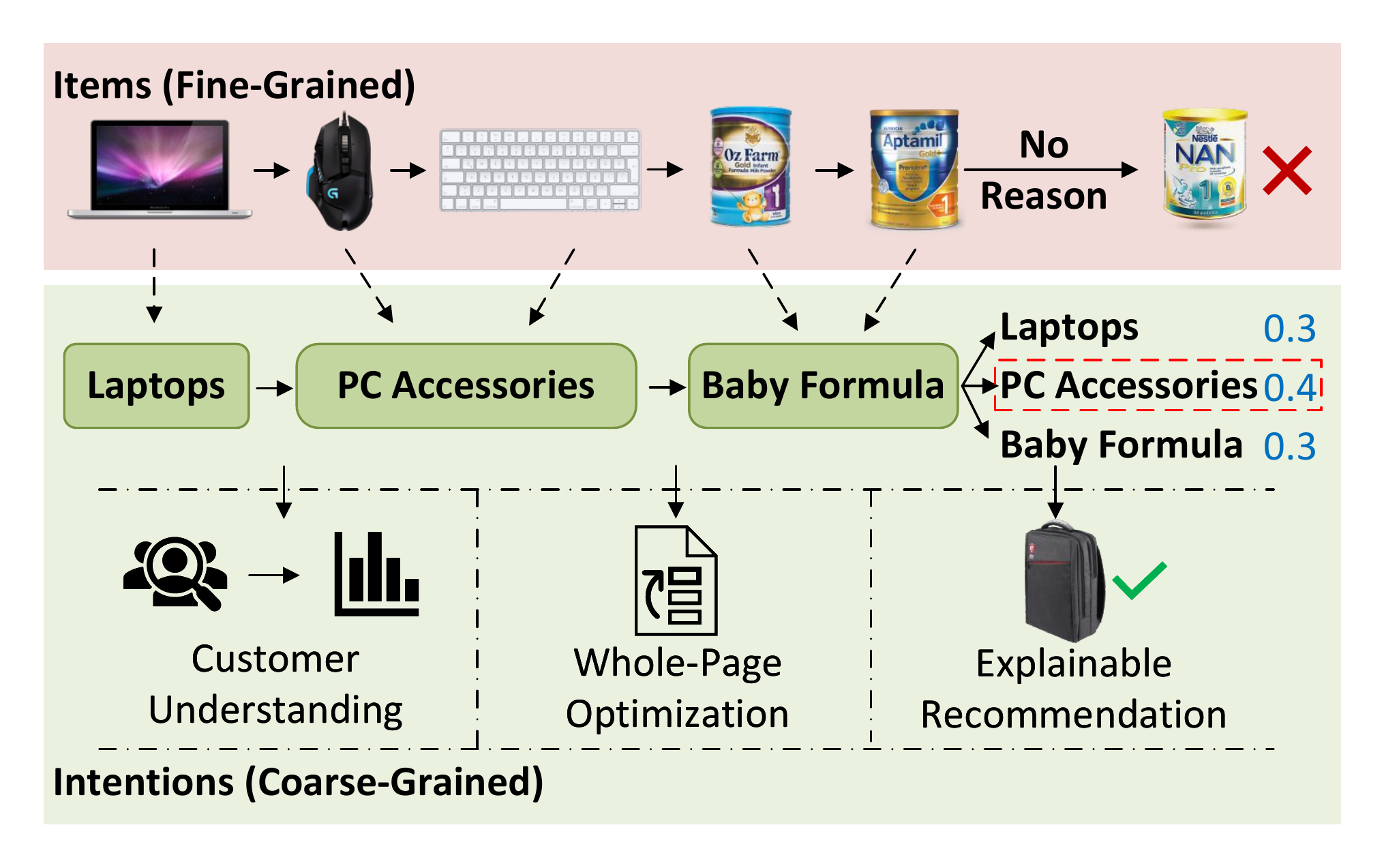}
\vspace{-3ex}
\caption{Illustration of sequential item recommendation and main shopping intention identification. For the majority of sequential recommendation algorithms, recommendations are provided without specific reasons, which can be explained by the main shopping intention identification.}
\vspace{-3ex}
\label{fig:intro}
\end{figure}

Sequential recommendation, which aims at understanding  evolving customer behaviors and dynamically recommending new items, has garnered considerable interests.
E-commerce stores, \eg, Amazon, have adopted  customers' \emph{shopping intentions} signals into  their recommendation systems~\cite{hao2020p}.
Although such initial works were built with a heuristic shopping intention detection approach, they already achieved significant improvement in terms of the \textit{MOI} metric (metric of interest), which considers both the short- and long-term effects on the customers' shopping experience.
Besides product-level recommendations, whole-page optimization, another downstream application, can also benefit from such customer intention signals.
Recent work~\cite{kanase2022An} has also  achieved significant lift of MOI on retail website homepage and checkout page through online A/B experimentation.
Thus, a robust, scalable, and explainable shopping intention detection approach plays crucial roles in multiple stages of the recommendation pipeline.

However, the majority of existing sequential recommendation algorithms~\cite{chen2018sequential, kang2018self, sun2019bert4rec, tang2018personalized} focus solely on product-level data to predict  subsequent recommendation items, regardless of the underlying shopping intentions.
This leaves these approaches inadequate in  capturing whole-sequence patterns, as they tend to place more emphasis on the item transitions learned from training data than on comprehending the customers' underlying objectives.
For example, when a user's interaction sequence is provided as in Figure~\ref{fig:intro}, most sequential recommendation systems will emphasize the commonly-seen transition pair, (\texttt{baby formula}$\to$\texttt{baby formula}), to recommend subsequent products without explanation.
In contrast, if we can identify the main shopping intention, we will find out that \texttt{PC Accessories} should be the most important intention for this customer and recommend products accordingly.

To this end, some approaches~\cite{li2022coarse, chang2021sequential} start to take both product-level and intention-level data into account, but merely use the concealed main shopping intentions as implicit guidance for the following item recommendation.
However, these implicit guidances can be severely influenced by the product-level interactions, making popular item transitions dominate the main intention detection.
Thus, explicitly identifying the customers' main shopping intentions to infer the user preferences becomes a prerequisite for explainable recommendations and user understanding.

To accomplish the main shopping intention identification task, the most typical and direct method is to map product-level interactions to intention-level sequences and then apply sequential recommendation algorithms.
Among the existing methods, recent advancements in Transformer~\cite{kang2018self, sun2019bert4rec, liu2021noninvasive} introduce the self-attention mechanism to reveal the position-wise item-item relations, which have achieved state-of-the-art performances.
Despite their success in product-level sequential recommendations, we argue that simply adapting such embedding-based Transformer architectures to intention-level sequences fail to incorporate:
(1) \textit{the shopping intention characteristics}: 
Shopping intentions are higher-level taxonomy, which can be considered as sets of products. 
Using only deterministic embeddings to represent shopping intentions is insufficient to capture this high-level characteristic;
(2) \textit{the user preferences composed of multiple intentions}: 
Users can have multiple intentions and preferences in mind during a shopping journey.
Using the Transformer architecture to characterize user preferences as deterministic points is also insufficient for estimating the relevance between user preferences and a composition of multiple shopping intentions;
(3) \textit{the collaborative transitivity}: 
Collaborative transitivity indicates the ability of introducing additional collaborative relevance beyond constrained intention transition pairs.
Transformer architectures employ dot-product-based attention mechanism, which is difficult to infer the relevance across pairs, (\eg, using \textit{a} and \textit{b}, \textit{b} and \textit{c} pairs to infer a and c are relevant as well);
(4) \textit{dynamic uncertainty}: In customer interaction sequences, it is common to witness a significant random shift in preferences without obvious correlations across intention transitions. Customers with more interests in dynamic variability are intuitively more uncertain. Therefore, when modeling user preferences, dynamic uncertainty is a vital component.


To this end, we present a new graph-regularized stochastic Transformer framework, \ours, for main shopping intention identification.
Our approach overcomes the aforementioned challenges with the following key designs:
(1) To better incorporate collaborative transitivity and uncertainty into representations, we describe each intention as an elliptical Gaussian distribution. 
Specifically, \ours 
applies stochastic embedding layers to assign each intention a mean and covariance embedding, composing the
stochastic representations; 
(2) To transfer knowledge from popular intentions to unpopular ones for cold-start issue resolution, we introduce global intention relation information as prior knowledge for improved intention modeling.
Specifically, we design an intention relation graph for regularization, with diverse intentions as nodes and complementary/relevant relations between intentions as edges.
By propagating intention representations on the graph, relevant shopping intentions are dragged towards each other on latent representation space to share close embedding distributions.
It can also alleviate the data scarcity issue for unpopular intentions, whose embeddings can be inferred from  their neighbors on the graph.
(3) Once we obtain the regularized stochastic representations for the users' interactions with intentions, we send them to mean and covariance Transformers to model the sequential information from intention transitions.
Instead of using dot product to compute relevance score between user preferences and recommendations in deterministic models, we apply Wasserstein distance to measure the distances between distributions.
Considering the distances as dissimilarity between intentions with uncertainty information, we combine it with Bayesian Personalize Ranking (BPR) loss~\cite{rendle2009bpr} as the training objective.

Our contributions can be summarized as follows:

    \noindent $\bullet$ To the best of our knowledge, this is the first work focusing on the main shopping intention identification task with solely intention-level data. This will help with user understanding and improve the performances of downstream tasks, including product-level recommendation, ranking stage of whole page optimization;
    
    \noindent $\bullet$  We describe the intentions as Gaussian distributions using stochastic representations to reflect the high-level properties of intentions, collaborative transitivity across intentions, and user preference uncertainty;
    
    \noindent $\bullet$  We introduce the shopping intention relation graph as prior knowledge and propose a novel graph regularizer to restrict the stochastic representations in distribution-based methods;
    
    \noindent $\bullet$  We develop three different Amazon real-world datasets, covering long-term, short-term, and purchase-related user cases. Our proposed \ours outperforms state-of-the-art baselines significantly by $18.08\%$ in Hit@1, $6.11\%$ in NDCG@10, and $7.01\%$ in Hit@10 on average of three datasets. 

\section{Related Works}
\subsection{Sequential Recommendation (SR)}\label{subsec:re-SR}
Sequential Recommendation (SR) aims to predict the next item based on the user historical interactions with products.
Earlier SR works, like FPMC~\cite{rendle2010factorizing} and Fossil~\cite{he2016fusing}, apply Markov Chains with matrix factorization to model the first-order and higher-order item-to-item transition matrices.
However, when encountering with long interaction sequences or considering long-term influence from previous items, the computations of modeling the transition matrices  increase exponentially.
More recent sequential recommendations can better learn  item-to-item transition patterns automatically
via deep learning architectures, including Convolution Neural Networks (CNN)~\cite{tang2018personalized}, Recurrent Neural Networks (RNN)~\cite{quadrana2017personalizing, devooght2017long, li2017neural, ma2019hierarchical, peng2021ham} and self-attentive models~\cite{kang2018self, sun2019bert4rec, wu2020sse}.
Among them, the Transformer-based models reach the state-of-the-art performances with the capability of extracting context information from all past actions and learn short-term dynamics.
However, they still struggle to solve the cold-start issue~\cite{yu2023coldstart}, which means for unpopular items, the representations are under-trained and it is difficult for models to make precise predictions over them without sufficient data.
In addition, despite the fact that these models have been successful in sequential recommendation systems and can be directly transferred to the main shopping intention identification task, they are incapable of capturing the unique characteristics of intentions, as intentions are higher-level concepts than items.

\subsection{Intention Identification in SR}\label{subsec:re-intent}
With the development of recommendation systems, people start to seek for other side information to help improve the recommendation qualities.
Shopping intentions usually serve as a coarse-grained side information that can better describe users' preferences.
Most of the existing methods~\cite{zhang2019feature, tanjim2020attentive, zhu2020sequential, liu2021noninvasive, li2022coarse, shen2023learning} focus on treating the intentions as an implicit guidance or an additional feature for downstream product-level recommendation.
However, the intention guidance learned from the sequences may not be in  line with expectation. As they are using the same model architecture on both intention and item sequences, the main shopping intention can be misled by some commonly seen yet totally irrelevant item pairs.
To resolve this issue, some other methods leverage clustering algorithms~\cite{chang2021sequential} or graph neural networks~\cite{wang2021learning, chen2022intent, chen2022global} to better understand the intentions via linking them with  items and users.
However, these approaches lose the explainability of shopping intentions for better user understanding.
Another  line of works identify the shopping intentions from search queries input by the users~\cite{hashemi2021learning, gong2021density, yang2021finn}, which is orthogonal to our work.

\subsection{Stochastic Representations in SR}\label{subsec:re-sto}
Representing concepts (\eg, natural language sequences, images, graphs) as distributions has attracted interests from the research community~\cite{kingma2013auto, bojchevski2018deep, he2015learning, qian2021conceptualized, shen2023semantic, shen2023fishrecgan}.
Most stochastic models represent the concepts with Gaussian distributions, composed of mean and covariance.
The distribution representations introduce uncertainties and provide more flexibility compared with deterministic embeddings. 
In the recommendation systems area, a few studies  propose to leverage the advantages of Gaussian distributions to  flexibly represent users and items.
For example, GeRec~\cite{jiang2019convolutional} models each user and item as a Gaussian distribution and applies CNN on the sampled matrix from their distributions for inference.
To dynamically monitor the sequential changes in user interactions, a series of work~\cite{zheng2019deep, fan2021modeling, fan2022sequential} also combine Gaussian distributions with sequential recommendation algorithms.
DT4SR~\cite{fan2021modeling} is one of these attempts, proposing the mean and covariance embeddings to model items as distributions.
STOSA~\cite{fan2022sequential} extends DT4SR architecture via proposing a new stochastic attention mechanism based on Wasserstein distance.
However, all the previous stochastic methods, learning the distributions only from the transitions within the sequence, can still fail in cold-start situations in \cref{subsec:re-SR}, which can be mitigated by \ours. 
In addition, none of the aforementioned methods naturally combine stochastic representations with intentions to learn the intention distributions, which is accomplished by \ours as well.
Another methodology line is variational autoencoder (VAE), approximating posterior distributions of latent variables via variational inference.
Combining SR and VAE, SVAE~\cite{sachdeva2019sequential} and VSAN~\cite{zhao2021variational} learn the dynamic hidden representations of shopping sequences.
However, these efforts still perform worse than existing deterministic models on many tasks~\cite{kang2018self, sun2019bert4rec}.
As an update, ACVAE~\cite{xie2021adversarial} incorporates adversarial variational Bayes and maximization of mutual information between user embeddings and input sequences to obtain more distinctive and personalized representations of individual users.
However, all these VAE-based SR methods are easy to suffer from posterior collapse problems, generating poor-quality latent representations, which \ours can avoid.


\section{Method}

\begin{figure*}[t]
\centering
\includegraphics[width=0.9\linewidth]{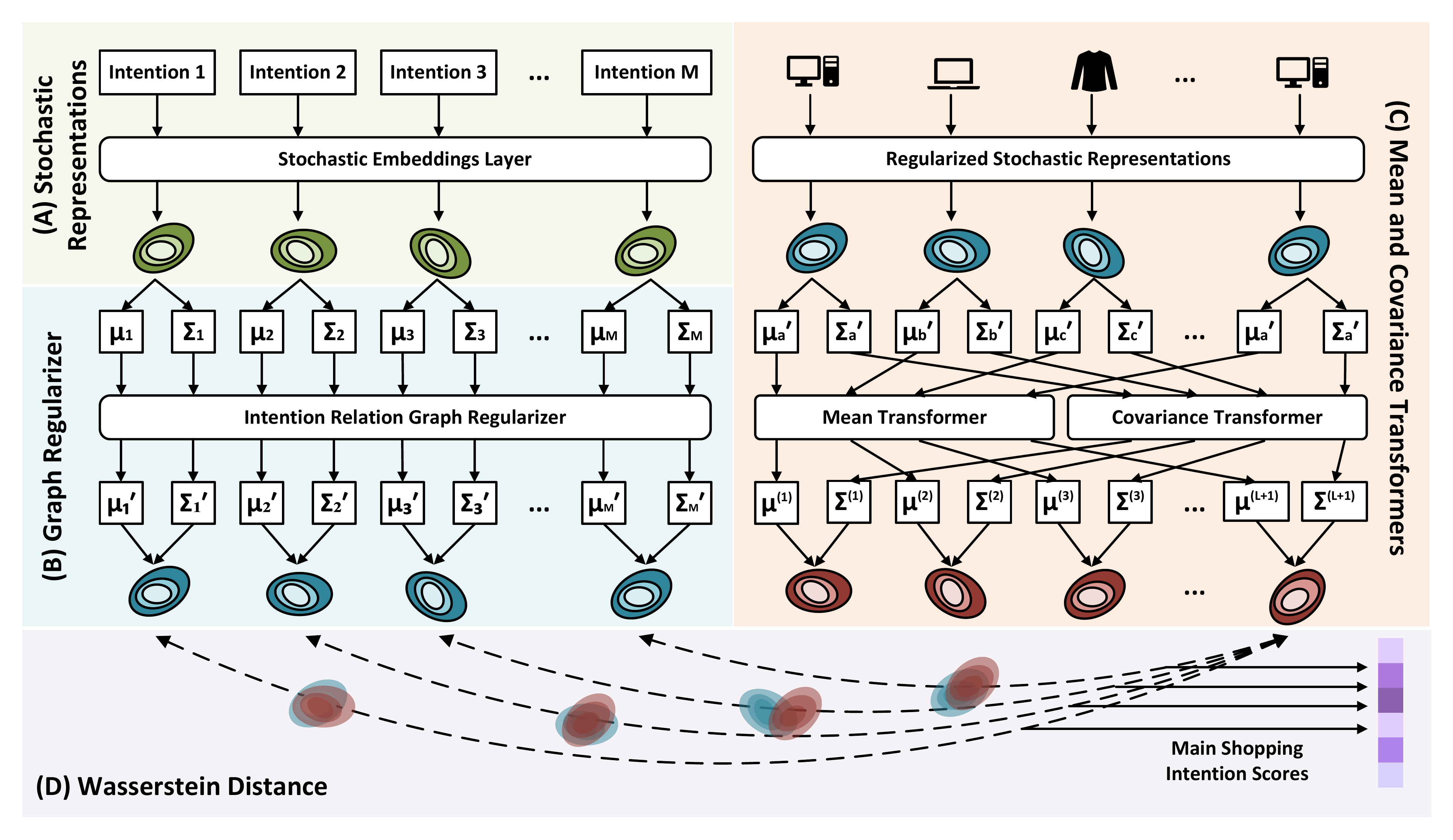}
\caption{Illustration of \ours model, containing key components: (A) stochastic representations to map each intention as a Gaussian distribution; (B) graph regularizer to restrict the more relevant intentions to have closer representations; (C) mean and covariance transformers to encode the sequential information from user historical interactions; (D) Wasserstein distance for training and inference.}
\vspace{-1ex}
\label{fig:model}
\end{figure*}

In this section, we introduce our proposed graph regularized stochastic model, \ours, for main shopping intention identification (\cref{prob}).
Figure~\ref{fig:model} displays the working flow of \ours.
It consists of several key components:
(1) stochastic representations (\cref{method:sel}), which model the shopping intentions as stochastic distributions, consisting of mean and covariance embeddings;
(2) intention relation regularizer (\cref{method:irgr}), which creates an intention relation graph and regularize more relevant intentions on the graph to have closer stochastic representations;
(3) mean and covariance Transformers (\cref{method:mct}), encoding sequential information from the transition patterns in user historical interactions to generate stochastic representations of user preferences;
(4) Wasserstein distance (\cref{method:te}), which measures the dissimilarity between intentions and user preferences and can be combined with Bayesian Personalized Ranking (BPR)~\cite{rendle2009bpr} loss on the positive and negative sequences for model training.

\subsection{Problem Definition}\label{prob}
A sequential recommendation system collects the interactions between a set of users $\mathcal{U}$ and items $\mathcal{V}$, (\eg, clicks, purchases, \etc) and sorts them chronologically into sequences.
Similarly, to identify users' main shopping intentions, models need to map all the items $\mathcal{V}$ to their belonging shopping intentions $\mathcal{M}$ and reorganize them chronically as $\mathcal{S}^u=[m_1^u, m_2^u, \cdots, m_{|\mathcal{S}^u|}^u]$.
The goal of the main shopping intention identification can be formulated as:
\begin{equation}
    \begin{aligned}
    p(m^u=m|\mathcal{S}^u),
    \end{aligned}
\end{equation}
which measures the probability of an intention $m$ being the main shopping intention $m^u$ given user $u$'s sequence.

\subsection{Stochastic Embedding Layers}\label{method:sel}
Different from the deterministic embedding layers that only map the items/intentions to unique high-dimensional vectors, 
stochastic embedding layers formulate the intentions as high-dimensional elliptical Gaussian distributions.
These Gaussian distributions are constructed of mean and covariance embeddings, spanning a broader space to include more high-dimensional points, which naturally captures high-level semantics of shopping intentions.
Specifically, we define a mean embedding table $\mathbf{T}^\mu\in\mathbb{R}^{|\mathcal{M}|\times d}$ and a covariance embedding table $\mathbf{T}^\Sigma\in\mathbb{R}^{|\mathcal{M}|\times d}$, where $|\mathcal{M}|$ denotes the total number of shopping intentions and $d$ denotes the hidden dimension size.
As the mean and covariance of a Gaussian distribution identify different signals of expectation and uncertainty, we introduce different position embeddings $\mathbf{P}^\mu\in\mathbb{R}^{L\times d}$ and $\mathbf{P}^\Sigma\in\mathbb{R}^{L\times d}$, where $L$ indicates the sequence length.
Thus, we can obtain the stochastic representations via the summation of the previously defined embeddings and position embeddings:
\begin{equation}\label{eq:311}
    \begin{aligned}
    \mathbf{E}^\mu_{\mathcal{S}^u}&=[\mathbf{E}_{s_1}^\mu, \cdots, \mathbf{E}_{s_L}^\mu]=[\mathbf{T}_{s_1}^\mu+\mathbf{P}_{s_1}^\mu, \cdots, \mathbf{T}_{s_L}^\mu+\mathbf{P}_{s_L}^\mu],\\
    \mathbf{E}^\Sigma_{\mathcal{S}^u}&=[\mathbf{E}_{s_1}^\Sigma, \cdots, \mathbf{E}_{s_L}^\Sigma]=[\mathbf{T}_{s_1}^\Sigma+\mathbf{P}_{s_1}^\Sigma, \cdots, \mathbf{T}_{s_L}^\Sigma+\mathbf{P}_{s_L}^\Sigma].
    \end{aligned}
\end{equation}
For example, the first intention $s_1$ in the sequence $\mathcal{S}^u$ is formulated as a Gaussian distribution $\mathcal{N}(\mu_{s_1},\Sigma_{s_1})$, where $\mu_{s_1}=\mathbf{E}_{s_1}^\mu$ and $\Sigma_{s_1}=\operatorname{diag}(\operatorname{elu}(\mathbf{E}_{s_1}^\Sigma)+\mathbf{1})\in\mathbb{R}^{d\times d}$.
$\operatorname{elu}(\cdot)$ is the ELU activation function and $\mathbf{1}\in\mathbb{R}^d$ is an all-ones vector.~\footnote{The operations on covariance embeddings are designed to guarantee the covariance matrix to be positive definite.}

\subsection{Intention Relation Graph Regularizer}\label{method:irgr}
To effectively model infrequent and under-trained shopping intentions, our goal is to utilize the most related intentions to help the model comprehend them.
To this end, we propose a novel graph-based regularizer, allowing more pertinent shopping intentions to share closer stochastic representations.
Thus, we introduce the global intention relationship as the prior knowledge and create a graph accordingly, which can be introduced as follows:

\noindent \textbf{Intention Relation Graph.}
We create the intention relation graph with the aid of P-Companion~\cite{hao2020p}.
Given a pair of shopping intentions $(m_i, m_j)$ as inputs, we treat the co-purchase relations between them as distant supervised labels $y_{i,j}=\{+1, -1\}$.
To extract  relevant information from co-purchase relations, each shopping intention and its complementary side are represented by two separate embeddings, $\phi_w,\phi_w^c$. 
Thus, to infer the relation between shopping intentions $(m_i, m_j)$, we first apply a 2-layer feed-forward network (FFN) to transform $m_i$:
\begin{equation}
    \begin{aligned}
    \gamma_{m_i}=\operatorname{ReLU}(\phi_{m_i}\mathbf{W}_1+\mathbf{b}_1)\mathbf{W}_2+\mathbf{b}_2,
    \end{aligned}
\end{equation}
where $\{\mathbf{W}_1,\mathbf{W}_2,\mathbf{b}_1,\mathbf{b}_2\}$ are trainable parameters. 
Then, we optimize the relation between the transformed embedding of $m_i$, $\gamma_{m_i}$, and the complementary embedding of $m_j$, $\phi_{m_j}^c$, via applying hinge loss function on labels $y_{i,j}$:
\begin{equation}
    \begin{aligned}
    \ell=\min\sum_{m_i,m_j\in\mathcal{M}}(\max\{0,\epsilon-y_{i,j}(\lambda-\|\gamma_{m_i}-\phi_{m_j}^c\|_2^2\}),
    \end{aligned}\label{eq:graphloss}
\end{equation}
where $\lambda$ is the base distance to distinguish $\gamma_{m_i}$ and $\phi_{m_j}^c$, and $\epsilon$ is the margin distance.
When $y_{i,j}=1$, the two shopping intentions have the co-purchase/complementary relation and the model will force the distance between $\gamma_{m_i}$ and $\phi_{m_j}^c$ to be smaller than $\lambda-\epsilon$.
Otherwise, when $y_{i,j}=-1$, the two shopping intentions do not possess the complementary relation, pushing the $\gamma_{m_i}$ and $\phi_{m_j}^c$ far away from each other with distance more than $\lambda+\epsilon$.
With the trained $\phi_{m_i}$ and $\phi_{m_i}^c$, we can create the shopping intention relation graph $\mathcal{G}=\{\mathcal{V},\mathcal{E}\}$, where the nodes $\mathcal{V}$ are different shopping intentions, and the edges $\mathcal{E}$ indicate the relevant/complementary scores between the intentions.
To compute edge weights between intentions $m_i$ and $m_j$, we apply cosine similarities between $m_i$'s complementary embedding, $\phi_{m_i}^c$, and $m_j$'s embedding, $\phi_{m_j}$:
\begin{equation}
    \begin{aligned}
    e(m_i,m_j)=\frac{\phi_{m_i}^c\cdot\phi_{m_j}}{\|\phi_{m_i}^c\|\|\phi_{m_j}\|}.
    \end{aligned}\label{eq:edge}
\end{equation}

\noindent \textbf{Graph Neural Regularizer.}
Considering the previously constructed relational graph of intentions as prior knowledge, we aim to regularize the more relevant intentions on the graph so that they share closer distributions.
When determining the main shopping intentions of a user, we can rank all relevant intentions higher, including those even unpopular ones, if they are represented similarly.
Thus, we employ the graph convolutional network (GCN) to induce stochastic representations of nodes based on the neighboring features, transferring knowledge to under-trained nodes from their frequently-seen neighbors.
To learn a unified set of parameters regularizing both the mean embeddings $\mathbf{T}^\mu\in\mathbb{R}^{|\mathcal{M}|\times d}$ and the covariance embeddings $\mathbf{T}^\Sigma\in\mathbb{R}^{|\mathcal{M}|\times d}$ simultaneously, we concatenate them together as the initial node representations:
\begin{equation}
    \begin{aligned}
    \mathbf{X}^{(0)}=\mathbf{T}^\mu\oplus\mathbf{T}^\Sigma\in\mathbb{R}^{|\mathcal{M}|\times 2d}.
    \end{aligned}
\end{equation}
Then, the GCN propagation can be represented as follows:
\begin{equation}
    \begin{aligned}
    \mathbf{X}^{(l)}=\sigma(\tilde{\mathbf{A}}\mathbf{X}^{(l-1)}\mathbf{W}^{(l-1)}),
    \end{aligned}
\end{equation}
where $\mathbf{W}^{(l-1)}\in\mathbb{R}^{2d\times 2d}$ is the trainable weight matrix after the $l$-th layer, and $\mathbf{X}^{(l)}\in\mathbb{R}^{|\mathcal{M}|\times 2d}$ indicates the obtained new regularized intention representations at the $l$-th layer. $\tilde{\mathbf{A}}\in\mathbb{R}^{|\mathcal{M}|\times|\mathcal{M}|}$ is the normalized format of the adjacent matrix $\mathbf{A}\in\mathbb{R}^{|\mathcal{M}|\times|\mathcal{M}|}$ derived from graph $\mathcal{G}$, which can be computed via $\tilde{\mathbf{A}}=\mathbf{D}^{-\frac{1}{2}}\mathbf{A}\mathbf{D}^{-\frac{1}{2}}$,
where $\mathbf{D}$ is the degree matrix~\footnote{$D_{ii}=\sum_jA_{ij}$.} of $\mathbf{A}$.
Once  obtain the regularized intention representations $\mathbf{X}^{(l)}\in\mathbb{R}^{|\mathcal{M}|\times 2d}$ at the $l$-th layer, we will separate the new mean embeddings $\hat{\mathbf{T}}^\mu\in\mathbb{R}^{|\mathcal{M}|\times d}$ and the new covariance embeddings $\hat{\mathbf{T}}^\Sigma\in\mathbb{R}^{|\mathcal{M}|\times d}$ from them to form the regularized Gaussian distributions with intention relations:
\begin{equation}
    \begin{aligned}
    [\hat{\mathbf{T}}^\mu\oplus\hat{\mathbf{T}}^\Sigma]=\mathbf{X}^{(l)}.
    \end{aligned}
\end{equation}
Thus, we can finally rewrite the embedding layer in Eq.(\ref{eq:311}) to encode the user shopping sequences with regularized distributions:
\begin{equation}
    \begin{aligned}
    \hat{\mathbf{E}}^\mu_{\mathcal{S}^u}&=[\hat{\mathbf{E}}_{s_1}^\mu, \cdots, \hat{\mathbf{E}}_{s_L}^\mu]=[\hat{\mathbf{T}}_{s_1}^\mu+\hat{\mathbf{P}}_{s_1}^\mu, \cdots, \hat{\mathbf{T}}_{s_L}^\mu+\hat{\mathbf{P}}_{s_L}^\mu],\\
    \hat{\mathbf{E}}^\Sigma_{\mathcal{S}^u}&=[\hat{\mathbf{E}}_{s_1}^\Sigma, \cdots, \hat{\mathbf{E}}_{s_L}^\Sigma]=[\hat{\mathbf{T}}_{s_1}^\Sigma+\hat{\mathbf{P}}_{s_1}^\Sigma, \cdots, \hat{\mathbf{T}}_{s_L}^\Sigma+\hat{\mathbf{P}}_{s_L}^\Sigma].
    \end{aligned}\label{eq:user}
\end{equation}

\subsection{Mean and Covariance Transformers}\label{method:mct}
Apart from the prior knowledge we obtain from the global intention relation information, we still need to encode the sequential information from the user historical interaction sequences.
Therefore, we propose mean and covariance Transformers to automatically learn the hidden patterns from the intention transitions in the sequences.
The deterministic Transformer-based models build up the self-attention mechanisms with the dot products between query $\mathbf{Q}$, key $\mathbf{K}$, and value $\mathbf{V}$.
In sequential recommendation, the query, key, and value are obtained from the linear transformations of the same sequence embedding $\hat{\mathbf{E}}_{\mathcal{S}^u}$.
However, in distribution-based stochastic models, we use mean and covariance embeddings to form a Gaussian distribution as intention and sequence representations.
Thus, we need two separate sets of $\mathbf{Q}$, $\mathbf{K}$, and $\mathbf{V}$ for both mean and covariance embeddings of the sequence:
\begin{equation}
    \begin{aligned}
    &\mathbf{Q}_\mu(\mathcal{S}^u)=\hat{\mathbf{E}}_{\mathcal{S}^u}^\mu\mathbf{W}_\mu^Q,\ \mathbf{K}_\mu(\mathcal{S}^u)=\hat{\mathbf{E}}_{\mathcal{S}^u}^\mu\mathbf{W}_\mu^K,\ \mathbf{V}_\mu(\mathcal{S}^u)=\hat{\mathbf{E}}_{\mathcal{S}^u}^\mu\mathbf{W}_\mu^V;\\
    &\mathbf{Q}_\Sigma(\mathcal{S}^u)=\hat{\mathbf{E}}_{\mathcal{S}^u}^\Sigma\mathbf{W}_\Sigma^Q,\ \mathbf{K}_\Sigma(\mathcal{S}^u)=\hat{\mathbf{E}}_{\mathcal{S}^u}^\Sigma\mathbf{W}_\Sigma^K,\ \mathbf{V}_\Sigma(\mathcal{S}^u)=\hat{\mathbf{E}}_{\mathcal{S}^u}^\Sigma\mathbf{W}_\Sigma^V;
    \end{aligned}
\end{equation}
where $\mathbf{W}_*^*\in\mathbb{R}^{d\times d}$ represent the learnable weight matrices in the linear transformation. Combining the computed query $\mathbf{Q}$, key $\mathbf{K}$, and value $\mathbf{V}$ with scaled dot-product attention. we can use the mean self-attention (MSA) and covariance self attention (CSA) to obtain newly generated sequence stochastic representations $\{\mathbf{z}_{\mathcal{S}^u}^\mu,\mathbf{z}_{\mathcal{S}^u}^\Sigma\}$:
\begin{equation}
    \begin{aligned}
    \mathbf{z}_{\mathcal{S}^u}^\mu&=\mathbf{MSA}(\mathcal{S}^u)=\sigma(\frac{\mathbf{Q}_\mu(\mathcal{S}^u)\mathbf{K}_\mu(\mathcal{S}^u)^\mathrm{T}}{\sqrt{d}})\mathbf{V}_\mu(\mathcal{S}^u);\\
    \mathbf{z}_{\mathcal{S}^u}^\Sigma&=\mathbf{CSA}(\mathcal{S}^u)=\sigma(\frac{\mathbf{Q}_\Sigma(\mathcal{S}^u)\mathbf{K}_\Sigma(\mathcal{S}^u)^\mathrm{T}}{\sqrt{d}})\mathbf{V}_\Sigma(\mathcal{S}^u).
    \end{aligned}
\end{equation}

In addition to uncovering sequential patterns from linear transformations, we leverage the feed-forward network (FFN) to endow the model with non-linearity.
The FFN with respect to both $\mathbf{MSA}$ and $\mathbf{CSA}$ at position $t$ are defined as:
\begin{equation}
    \begin{aligned}
    \mathbf{F}_\mu(\mathcal{S}_t^u)&=\operatorname{FFN}^\mu(\mathbf{MSA}(\mathcal{S}_t^u))=\operatorname{elu}(\mathbf{z}_{\mathcal{S}_t^u}^\mu\mathbf{W}_1^\mu+\mathbf{b}_1^\mu)\mathbf{W}_2^\mu+\mathbf{b}_2^\mu;\\
    \mathbf{F}_\Sigma(\mathcal{S}_t^u)&=\operatorname{FFN}^\Sigma(\mathbf{CSA}(\mathcal{S}_t^u))=\operatorname{elu}(\mathbf{z}_{\mathcal{S}_t^u}^\Sigma\mathbf{W}_1^\Sigma+\mathbf{b}_1^\Sigma)\mathbf{W}_2^\Sigma+\mathbf{b}_2^\Sigma;
    \end{aligned}
\end{equation}
where all the $\mathbf{W}_*^*\in\mathbb{R}^{d\times d}$ and $\mathbf{b}_*^*\in\mathbb{R}^d$ are trainable parameters in feed-forward networks.


\subsection{Training and Evaluation}\label{method:te}
\noindent \textbf{Wasserstein Distance.}
For embedding-based models, when measuring how accurately the model detects the main intention, we need to apply dot products between the user preference embedding and the intention embeddings.
Similarly, for stochastic models, we need to identify the distances between distributions of the ground-truth labels and the inferred distributions. 
Many existing works formulating concepts as distributions use Kullback-Leibler (KL) divergence to compute the distribution distances.
However, when two intentions are excessively unrelated, their stochastic representations will be expressed as two almost non-overlapping distributions, and the KL divergence will describe the distance as nearly infinity, resulting in value instability.
Thus, we use Wasserstein distance to measure the distance between Gaussian distributions.
Given two Gaussian distributions $q_i=\mathcal{N}(\mu_i,\Sigma_i)$ and $q_j=\mathcal{N}(\mu_j,\Sigma_j)$, the Wasserstein distance can be computed as:
\begin{equation}
    \begin{aligned}
    d_W(i,j)=\|\mu_i-\mu_j\|_2^2+\operatorname{trace}(\Sigma_i+\Sigma_j-2(\Sigma_j^{1/2}\Sigma_i\Sigma_j^{1/2})^{1/2}).
    \end{aligned}
\end{equation}
For time efficiency, the second term in the above equation can be simplified as a calculation of Euclidean norm:
\begin{equation}
    \begin{aligned}
    \operatorname{trace}(\Sigma_i+\Sigma_j-2(\Sigma_j^{1/2}\Sigma_i\Sigma_j^{1/2})^{1/2})=\|\Sigma_i^{1/2}-\Sigma_j^{1/2}\|_F^2,
    \end{aligned}
\end{equation}
where $\|\cdot\|_F^2$ is Frobenius norm that can be calculated by matrix multiplications.

\noindent \textbf{Training Objective.}
Many deterministic sequential recommendation models apply Bayesian Personalize Ranking (BPR) loss~\cite{rendle2009bpr} on the dot-product scores, making the ground-truth label nearest to the customer preference.
In \ours, we apply BPR loss on the previous defined Wasserstein distances between distributions to measure the correctness of the main intention identification:
\begin{equation}
    \begin{aligned}
    \ell=-\sum_{\mathcal{S}^u\in\mathcal{S}}\sum_{t\in\{1,2,\cdots,|\mathcal{S}^u|\}}\log(\sigma(d_W(m_t^+, \hat{p}_t)-d_W(m_t^-,\hat{p}_t))),
    \end{aligned}\label{eq:loss}
\end{equation}
where $\hat{p}_t$ dentoes the inferred distribution of user preference at position $t$, $m_t^+$ is the ground-truth shopping intention and $m_t^-$ denotes the negative samples from the intentions that the user never interact with.
During the inference stage, for user $u$, we calculate the distances between the customer preference distribution and the candidates set containing the ground-truth intention and 100 negative sampled intentions.

Algorithm~\ref{alg:full} shows the whole working flow of \ours.

\begin{algorithm}[htb]
	\begin{small}
	\KwIn{$\mathcal{S}_u$: User Shopping Sequence.}
	\blue{// Step 1: \textit{Construct the Intention Relation Graph.}} \\ 
    Train the co-purchase relation model via Eq.~\ref{eq:graphloss}.\\
    Considering intentions as nodes, compute edge weights via Eq.~\ref{eq:edge}.\\
   \blue{// {Step 2}: \textit{Compute Graph-Regularized Embeddings.}} \\
    Use embedding layer to map intentions into mean and covariance embeddings: $\mathbf{T}^\mu$ and $\mathbf{T}^\Sigma$.\\
    Regularize $\mathbf{T}^\mu$ and $\mathbf{T}^\Sigma$ with intention relation graph as: $\hat{\mathbf{T}}^\mu$ and $\hat{\mathbf{T}}^\Sigma$.\\
    Map the user shopping sequences $\mathcal{S}^u$ into mean and covariance embedding sequences: $\hat{\mathbf{E}}^\mu_{\mathcal{S}^u}$ and $\hat{\mathbf{E}}^\Sigma_{\mathcal{S}^u}$ via Eq.\ref{eq:user}.\\
 \blue{// Step 3: \textit{Encode Sequential Information.}} \\
Use Mean and Covariance Transformers to model user preferences $\mathbf{F}_{\mu}(\mathcal{S}^u_t)$ and $\mathbf{F}_{\Sigma}(\mathcal{S}^u_t)$.\\
Update the model parameters via Eq.~\ref{eq:loss}.\\
	\KwOut{The final user main shopping intentions for the whole shopping trajectory.}
	\end{small}
	\caption{Process of \ours. }
	\label{alg:full}
\end{algorithm}

\section{Experiments}
In this section, we evaluate the empirical effectiveness \ours 
by studying the following research questions (RQs):

\noindent $\bullet$\textbf{RQ1:} Does \ours provide better shopping intention identification results than baselines?

\noindent $\bullet$\textbf{RQ2:} Why do we need to design different kinds of scenarios? How does \ours perform on different circumstances?

\noindent $\bullet$\textbf{RQ3:} What is the influence of the intention relation graph regularizer and stochastic representations?

\noindent $\bullet$\textbf{RQ4:} How sensitivity is \ours to the hyper-parameters?

\noindent $\bullet$\textbf{RQ5:} Why can \ours alleviate intention cold start issue?
\begin{table*}[htb]
\fontsize{8}{10}\selectfont \setlength{\tabcolsep}{0.3em}
\centering
\vspace{-1ex}
\caption{An example of generated sequences under different scenarios. Different numbers represent different shopping intentions. The \textcolor{blue}{blue} intentions indicate the validation set labels and the \textcolor{red}{red} intentions indicate the test set labels.}\label{tab:seq}
\setlength{\tabcolsep}{1.2mm}{
\begin{tabular}{@{}p{0.08\linewidth}p{0.4\linewidth}p{0.45\linewidth}@{}}
\toprule
\textbf{Scenarios} & \textbf{Examples}                                                    & \textbf{Explanation}                                               \\ \midrule
Original  & \big (0, 4293, 234, 2173, 232, 183, 913, 4298, 582, \textcolor{blue}{98}, \textcolor{red}{4299}\big ) & Raw Sequences obtained from user historical data; \\ \midrule
24-Hours  & \big (0, 4293, 234, 2173, \textcolor{blue}{232}, \textcolor{red}{183}\big), \quad \big (913, 4298, 582, \textcolor{blue}{98}, \textcolor{red}{4299}\big ) & Time interval between 183 and 913 is longer than 24 hours; \\ \midrule
\multirow{2}{*}{Purchase}  & \big (377, 19, 76, 6, 87, \textcolor{blue}{112}\big )\sout{, 90, 4, 346, 85} & 112 is the PURCHASE action.                                \\ 
& \big (0, 4293, 234, 2173, 232, \textcolor{red}{183}\big )\sout{, 913, 4298, 582, 98, 4299} & 183 is the PURCHASE action;\\ \bottomrule
\end{tabular}}
\vspace{-2ex}
\end{table*}

\subsection{Data Curation}
We create three benchmarks for main shopping intention identification using anonymized data from \textit{amazon.com}.
To provide a comprehensive evaluation, the ground-truth main shopping intention labels of the three datasets are created based on different real-life scenarios:
(1) \textit{original sequences}, using the raw sequences composed of user historical interactions to model the long-term shopping scenario;
(2) \textit{24-hours sequences}, leveraging more frequent user-intention interactions sampled from raw data to model the short-term shopping scenario;
(3) \textit{purchase sequences}, considering purchase actions of the raw sequences as strong signal of the main shopping intentions to model the purchase-related shopping scenario.
Table~\ref{tab:seq} shows examples of the three datasets, and the statistics of three datasets, original, 24-hours, purchase sequences, are listed in Table.~\ref{tab:dataset-appendix}.

\begin{table}[t]
\caption{Statistics of original, 24-hours, and purchase data.}\label{tab:dataset-appendix}
\centering
\footnotesize
\begin{tabular}{@{}lccc@{}}
\toprule
         & \multicolumn{1}{l}{\#Users} & \multicolumn{1}{l}{\#Intentions} & \multicolumn{1}{l}{Average \#Interactions/Seq} \\ \midrule
Original & 140000                      & 14695                            & 49.28                                          \\
24-hours & 130000                      & 14695                            & 15.66                                          \\
Purchase & 90000                       & 14695                            & 38.69                                          \\ \bottomrule
\end{tabular}
\end{table}

\noindent \textbf{Original sequences:} We follow the "leave-one-out" strategy~\cite{kang2018self} to split the sequences into training, validation, and test datasets.
To create the intention labels for each user and each split, we partition the curated historical sequence $\mathcal{S}^u$ for each user $u$ into three parts: (1) the most recent intention action $\mathcal{S}^u_{|\mathcal{S}^u|}$ as the intention label for test set; (2) the second most recent action $\mathcal{S}^u_{|\mathcal{S}^u|-1}$ as the intention label for validation set; and (3) all remaining actions as training data. Note that during testing, the input sequences contain training actions and validation action.

\noindent \textbf{24-hours sequences:} Although the original sequences can more accurately describe the long-term preferences of customers and be more useful for the downstream task of next item recommendation, the users' final interactions cannot always convey the main shopping intention labels due to the random shift of interests.
Assuming that, for more dense and frequent interactions in a short period, the last intentions from "leave-one-out" mechanism can better reflect the main shopping intentions, we assess the time intervals between the successive activities to breakdown the raw sequences.
If there is a temporal gap longer than a pre-determined threshold (\eg, 24 hours), we will insert a break-point and divide the entire sequence into two sub-sequences.

\noindent \textbf{Purchase sequences:} Aside from using the last intentions as the labels, the purchase actions can also serve as a very strong positive signal for main shopping intention identification. For each customer's historical data, the purchase actions can be viewed as the main shopping intentions for its preceding sub-sequences.
As purchasing activities are dispersed over the sequences, we can only apply user-based split for train/val/test separation in this scenario.


\subsection{Experiment Setup}

\subsubsection{Evaluation Protocol}
We evaluate all models with the following metrics:
(1) \textbf{NDCG@10:} A position-aware metric which assigns larger weights for higher positions;
(2) \textbf{Hit@K, (K=1,2,5,10):} Metrics counting the fraction of times that the ground-truth intention is among top $K$ predictions.
We report the averaged metrics over all users.
We select the model to report the test set performances based on the best validation NDCG@10 score.

\subsubsection{Baselines}
We compare our proposed model with baselines from three different groups:
(1) \textit{static recommendation methods}:
\textbf{Count-based Bayesian (CB)} is a non-learning approach that solely considers the appearance frequency of shopping intentions. We consider the intention frequencies across the entire market as prior, and the intention frequencies in each user's shopping sequence as likelihood. The final shopping intention rankings are derived using posterior, which is the multiplication of prior and likelihood;
\textbf{LightGCN~\cite{he2020lightgcn}} is a state-of-the-art graph-based static recommendation method, which considers high-order collaborative signals in user-item graph;
(2) \textit{deterministic sequential recommendation methods}:
\textbf{SASRec~\cite{kang2018self}} is a self-attention based sequential recommendation system model that captures long-term semantics and short-term dynamics;
(3) \textit{stochastic sequential recommendation methods}:
\textbf{DT4SR~\cite{fan2021modeling}} is a distribution-based method, mapping the intentions to elliptical Gaussian distributions and then send them into two separate Transformer-based model to infer the users' preferences;
\textbf{STOSA~\cite{fan2022sequential}} is a state-of-the-art distribution-based recommendation system model. STOSA extends DT4SR by proposing a new stochastic self-attention mechanism to further improve the combination of Transformer and stochastic representations;
(4) \textit{VAE-based sequential recommendation methods:}
\textbf{SVAE~\cite{sachdeva2019sequential}} is a recurrent version of VAE, combining recurrent neural network (RNN) and VAE. The model outputs the probability distribution of the most likely future preferences at each time step;
\textbf{ACVAE~\cite{xie2021adversarial}} is a state-of-the-art VAE-based model, first introducing the adversarial training for sequence generation, enabling the model to generate high-quality latent variables.

\subsubsection{Implementation Details}
For the hyper-parameters, the learning rate is set as $1e-4$, the maximum number of epochs is 500, the batch size is $128$. 
For the stochastic representations, the hidden dimensionalities of mean and covariance embeddings are set as $64$. 
For the intention relation graph regularizer, we apply 1-layer graph convolution network (GCN).
We train and test our code on the system Ubuntu 18.04.4 LTS with CPU: Intel(R) Xeon(R) Silver 4214 CPU@ 2.20GHz and GPU: NVIDIA V100.
We implement our method using Python 3.8 and PyTorch 1.6~\cite{paszke2019pytorch}.
For the hyper-parameters, the learning rate is set as $1e-4$, the maximum number of epochs is 500, the batch size is $128$. 
For the stochastic representations, the hidden dimensionalities of mean and covariance embeddings are set as $64$. 
For the intention relation graph regularizer, we apply 1-layer graph convolution network (GCN).
During training, we use the Adam~\cite{kingma2014adam} optimizer with $\beta_1=0.9$ and $\beta_2=0.999$ in our experiments for all the models. We select the best set of hyper-parameters of the models based on the NDCG@10 on the corresponding validation sets.
\subsection{Performance Comparison}

\begin{table*}[]
\centering
\fontsize{7.5}{9.5}\selectfont \setlength{\tabcolsep}{0.2em}
\caption{Performance Comparison in Hit@1, NDCG@10, Hit@2, Hit@5, and Hit@10 on three different datasets. The best results are boldfaced.}\label{tab:main}
\setlength{\tabcolsep}{1.2mm}{
\begin{tabular}{@{}lccccc|ccccc|ccccc@{}}
\toprule
Scenarios ($\rightarrow$) & \multicolumn{5}{c|}{Original}              & \multicolumn{5}{c|}{24-Hours}              & \multicolumn{5}{c}{Purchase}               \\ \cmidrule(l){2-16} 
Methods($\downarrow$)   & Hit@1 & NDCG@10 & Hit@2 & Hit@5 & Hit@10 & Hit@1 & NDCG@10 & Hit@2 & Hit@5 & Hit@10 & Hit@1 & NDCG@10 & Hit@2 & Hit@5 & Hit@10  \\ \midrule
\multicolumn{16}{l}{\emph{Static Recommendation}}  \\\midrule 
CB & 0.3242 & 0.4007 & 0.5224 & 0.7268 & 0.7708 & 0.3327 & 0.3949 & 0.5235 & 0.7233 & 0.7542 & 0.2767 & 0.3738 & 0.4376 & 0.6753 & 0.7502  \\
LightGCN & 0.2810 & 0.3528 & 0.4743 & 0.6461 & 0.7436 & 0.3545 & 0.4423 & 0.5411 & 0.7265 & 0.8783 & 0.2281 & 0.3635 & 0.4054 & 0.6470 & 0.7038\\\midrule
\multicolumn{16}{l}{\emph{Deterministic Sequential Recommendation}}  \\\midrule 
SASRec & 0.3192 & 0.3886 & 0.4509 & 0.6529 & 0.7865 & 0.4341 & 0.4537 & 0.5824 & 0.7610 & 0.8708 & 0.2761 & 0.3763 & 0.4160 & 0.6372 & 0.7804\\\midrule
\multicolumn{16}{l}{\emph{Stochastic Sequential Recommendation}}  \\\midrule
DT4SR & 0.4879 & 0.4417 & 0.5908 & 0.7231 & 0.8197 & 0.5950 & 0.5003 & 0.7027 & 0.8245 & 0.8968 & 0.4410 & 0.3863 & 0.5202 & 0.6308 & 0.7097 \\
STOSA & 0.5646 & 0.4736 & 0.6598 & 0.7714 & 0.8530 & 0.6240 & 0.5484 & 0.7106 & 0.8298 & 0.9183 & 0.6017 & 0.4781 & 0.6779 & 0.7763 & 0.8449\\\midrule
\multicolumn{16}{l}{\emph{VAE-based Sequential Recommendation}}  \\\midrule
SVAE & 0.2722		& 0.2666	&	0.3014	&	0.3573	&	0.4041 & 0.3142 & 0.3718 & 0.4863 & 0.7163 & 0.8257 & 0.2897 & 0.3706 & 0.4221 & 0.6419 & 0.7871\\ 
ACVAE & 0.3866		& 0.4582	&	0.4843	&	0.6712	&	0.7107 & 0.3264 & 0.4573  & 0.5286 & 0.7261 & 0.9125 & 0.2858 & 0.3743 & 0.4834 & 0.6682 & 0.7644\\ \midrule
\textbf{Ours} & \textbf{0.7061} & \textbf{0.5297} & \textbf{0.7890} & \textbf{0.8623} & \textbf{0.9050} & \textbf{0.8985} & \textbf{0.6115} & \textbf{0.9592} & \textbf{0.9896} & \textbf{0.9955} & \textbf{0.7280} & \textbf{0.5423} & \textbf{0.8062} & \textbf{0.8818} & \textbf{0.9259}\\ \bottomrule
\end{tabular}}
\end{table*}

\noindent \textbf{Cross-Method Comparison (RQ1).} 
Table~\ref{tab:main} reports the performances of \ours and the baselines on all three benchmarks.
The results demonstrate that \ours consistently outperforms all baselines in terms of all metrics on all the three datasets. 
Compared with the strongest baseline, STOSA, \ours shows significant improvements of $18.08\%$ of Hit@1, $6.11\%$ of NDCG@10, $16.87\%$ of Hit@2, $11.87\%$ of Hit@5, and $7.01\%$ of Hit@10 on average of three datasets.
From the results, we have the following observations:

\noindent (1) The performance gaps between our method and static methods, CB and LightGCN, show the importance of temporal order sequential information.
Different from product-level recommendations, the intentions may appear multiple times in single sequence, making counts-based method, CB, achieve comparable results to state-of-the-art static method of LightGCN, and even SASRec.

\noindent (2) Comparing our method with the backbone model, SASRec, \ours shows significant improvement of $43.44\%$ in Hit@1 and $15.50\%$ in NDCG@10.
This indicates that the stochastic representations expand the latent space for user-intention interactions and equip the model with collaborative transitivity. Besides, the graph regularizer captures global intention information, enabling model to better understand the intentions.
Both modules help enhance the intention identification capabilities, particularly in small-data  situations;

\noindent (3) The comparison between \ours and other distribution-based recommendation methods, DT4SR and STOSA, shows the efficacy of leveraging the intention relation graph as prior knowledge to regularize the stochastic representations and reveals the potential to further incorporate \ours with distribution-based attention;

\noindent (4) Comparing \ours with VAE-based methods, we find that VAE-based methods are easy to suffer posterior collapse problems: 
if the decoder is too expressive, the KL divergence term in the loss will converge to 0, generating similar latent representations for all inputs. 
This situation becomes worse for less-interacted under-trained intentions. 
On the contrary, our model leverages graph regularization to transfer knowledge from more-interacted intentions to less-interacted ones to mitigate this issue. 

\begin{figure}[tb]
\centering
\includegraphics[width=0.9\linewidth]{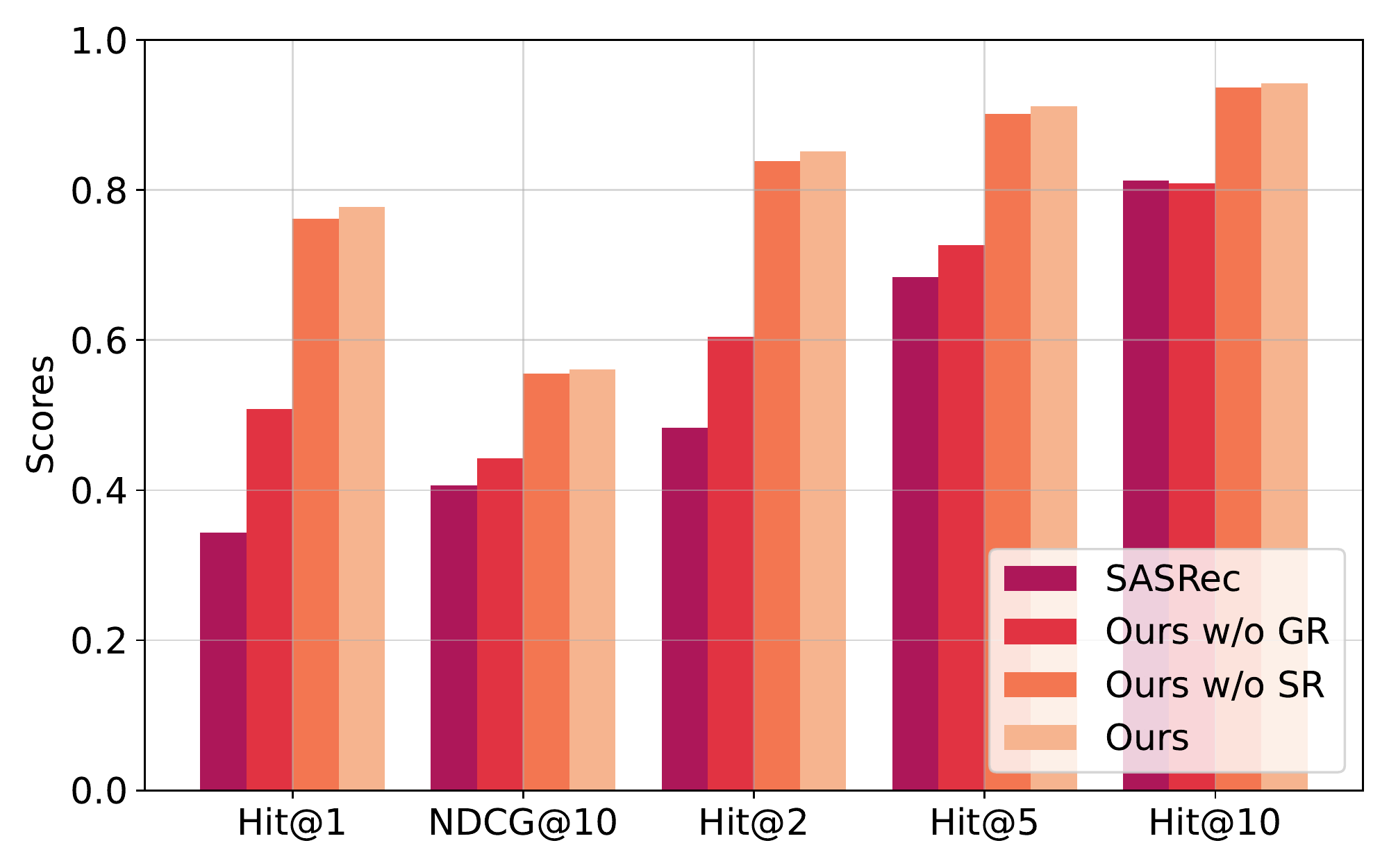}
\caption{Ablation Study in Hit@1, NDCG@10, Hit@2, Hit@5, Hit@10. The performances are averaged on three datasets. }
\vspace{-3ex}
\label{fig:abla}
\end{figure}

\noindent \textbf{Cross-Dataset Comparison (RQ2).} From Table~\ref{tab:main}, we can also compare the performances horizontally to gain insights into the models’ performances in different scenarios. 
We notice that the performance improvements made by \ours, compared with baselines, varies by different scenarios.
The performance gap is larger on 24-hours and purchase sequences than on original sequences.
We summarize the reasons as follows:
(1) On 24-hour sequences, \ours and all the baselines achieve higher absolute performances than on the other two categories.
As the 24-hour sequences consist of more dense and frequent activities, where users' severe random interest shifts are less likely to appear, and the hidden sequential dynamic patterns are easier for models to learn and capture.
(2) On purchase sequences, \ours performs slightly better, while most baselines perform poorly. 
This is because of the difference of data split for train/validation/test set:
we apply user-based split on purchase sequences, and "leave-one-out" on the other two categories.
Thus, it is easier for model to face new users/intentions during validation or testing, making the models more likely to suffer from the cold-start issue.
With the graph-regularized stochastic Transformer, \ours can better resolve this issue than the other baselines, enlarging the performance gap on this dataset.


\subsection{A/B Test}
We have an existing products recommendation feature using a static mapping created from the work~\cite{hao2020p}. However, the procedure only provides static product to product recommendation, it doesn't take customer's shopping history and provide personalized shopping experience. 
Our goal was to improve the recommendation's relevance by taking customers’ shopping intentions into consideration while optimizing sales and revenue. 
Our work, G-STO, has been recently deployed for online A/B testing as the treatment group, while the control group is the existing solution using static mapping. The model takes customer's past shopping activities as input, then predict the possible product categories for downstream recommendations. This online testing has been run for 2 weeks, and we have already found the achievements shown below:

We observe a World Wide (WW) commercial success:
\begin{itemize}
\item WW annual revenue was improved by 1\%;
\item WW annual sales were improved by 2\%.
\end{itemize}
\begin{figure}[tb]
\centering
\includegraphics[width=0.9\linewidth]{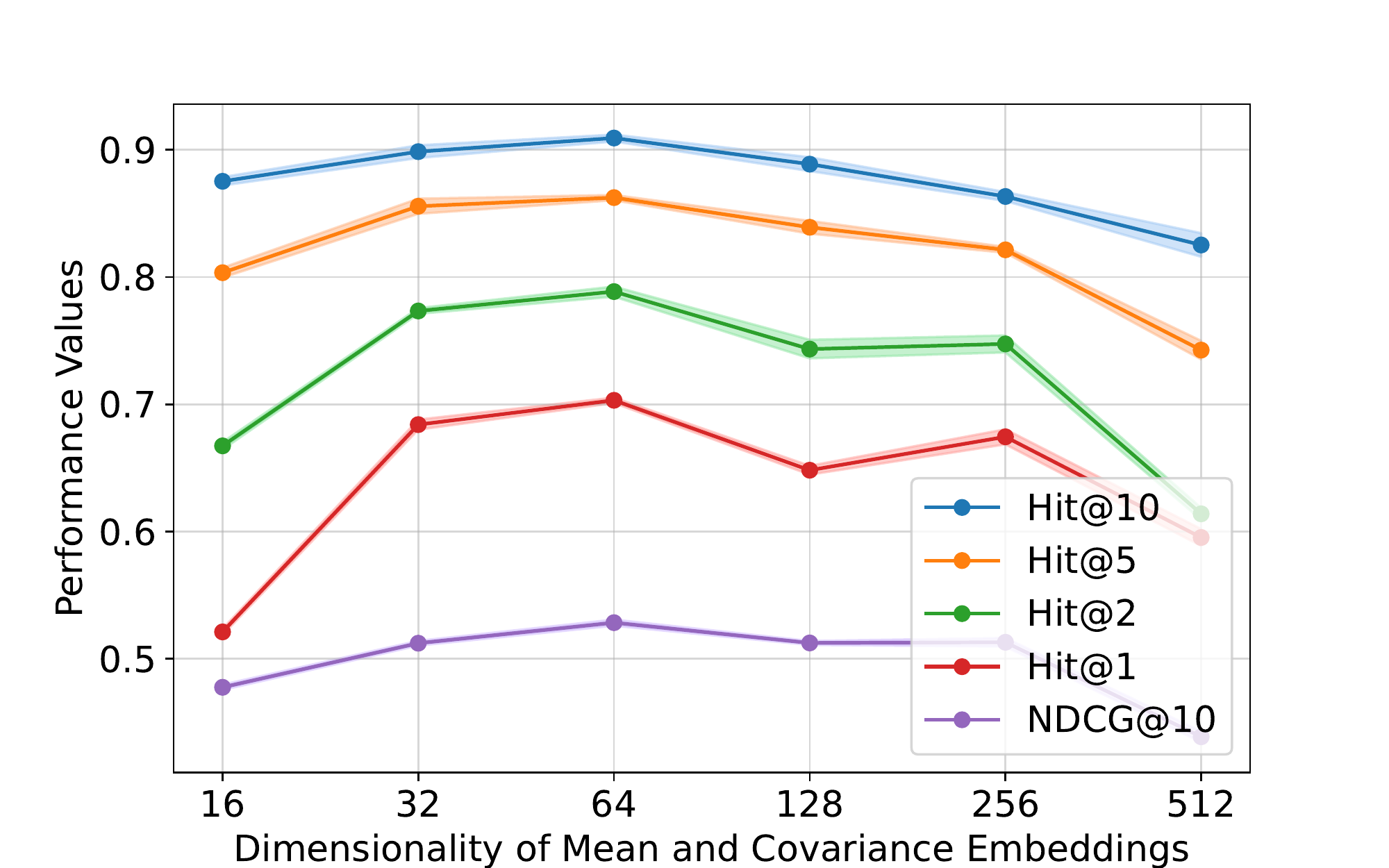}
\caption{Parameter studies of our model about the hidden dimensionality of mean and covariance embeddings on original sequences.The red line indicates that the best performances are obtained when the hidden size is 64. }
\vspace{-2ex}
\label{fig:para}
\end{figure}

\begin{figure*}[!ht]
\centering
\subfigure[Intention embeddings trained by SASRec.]{
\label{fig:tsne-sasrec}
\includegraphics[width=0.3\linewidth]{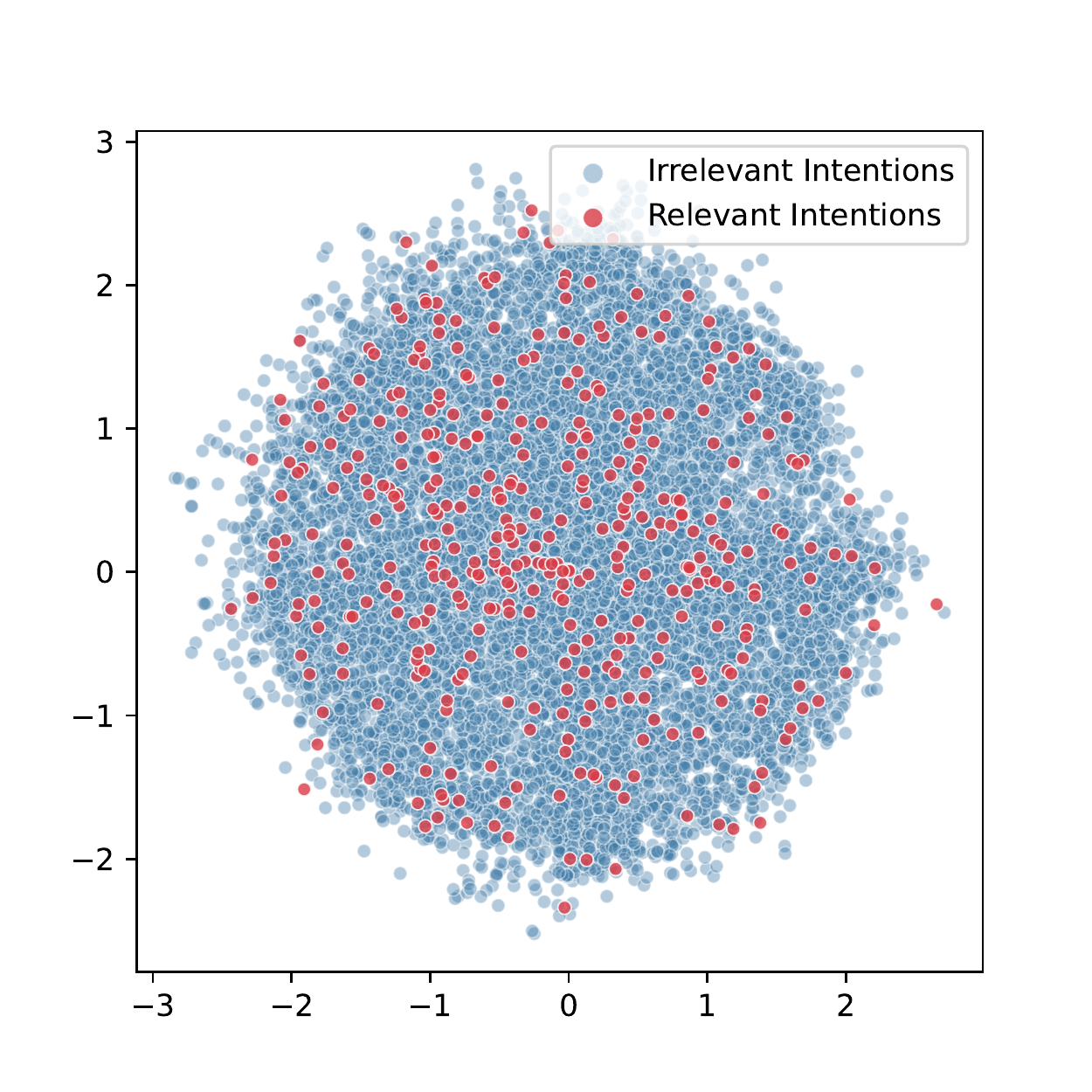}}
\subfigure[Intention embeddings trained by STOSA.]{
\label{fig:tsne-stosa}
\includegraphics[width=0.3\linewidth]{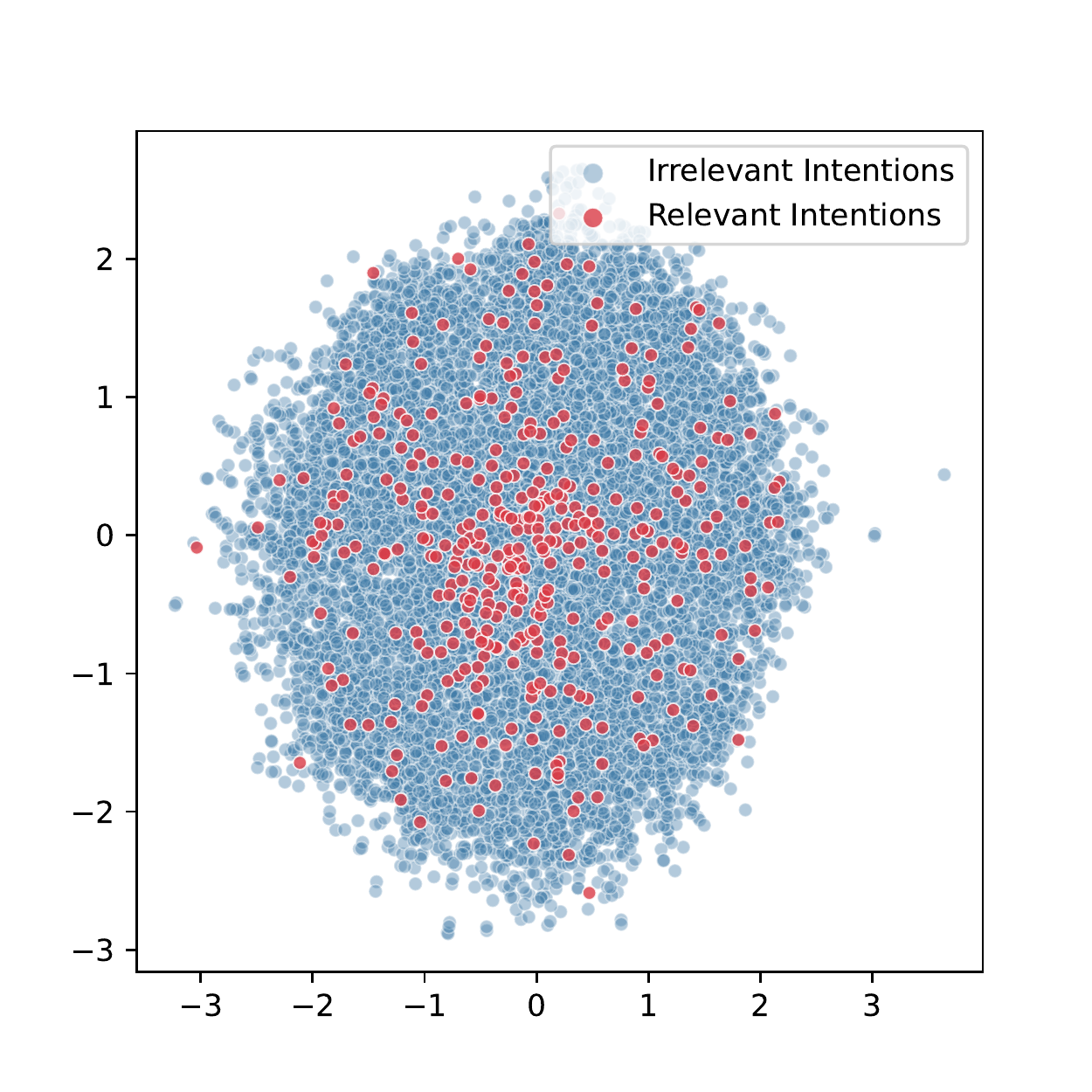}}
\subfigure[Intention embeddings trained by \ours.]{
\label{fig:tsne-ours}
\includegraphics[width=0.3\linewidth]{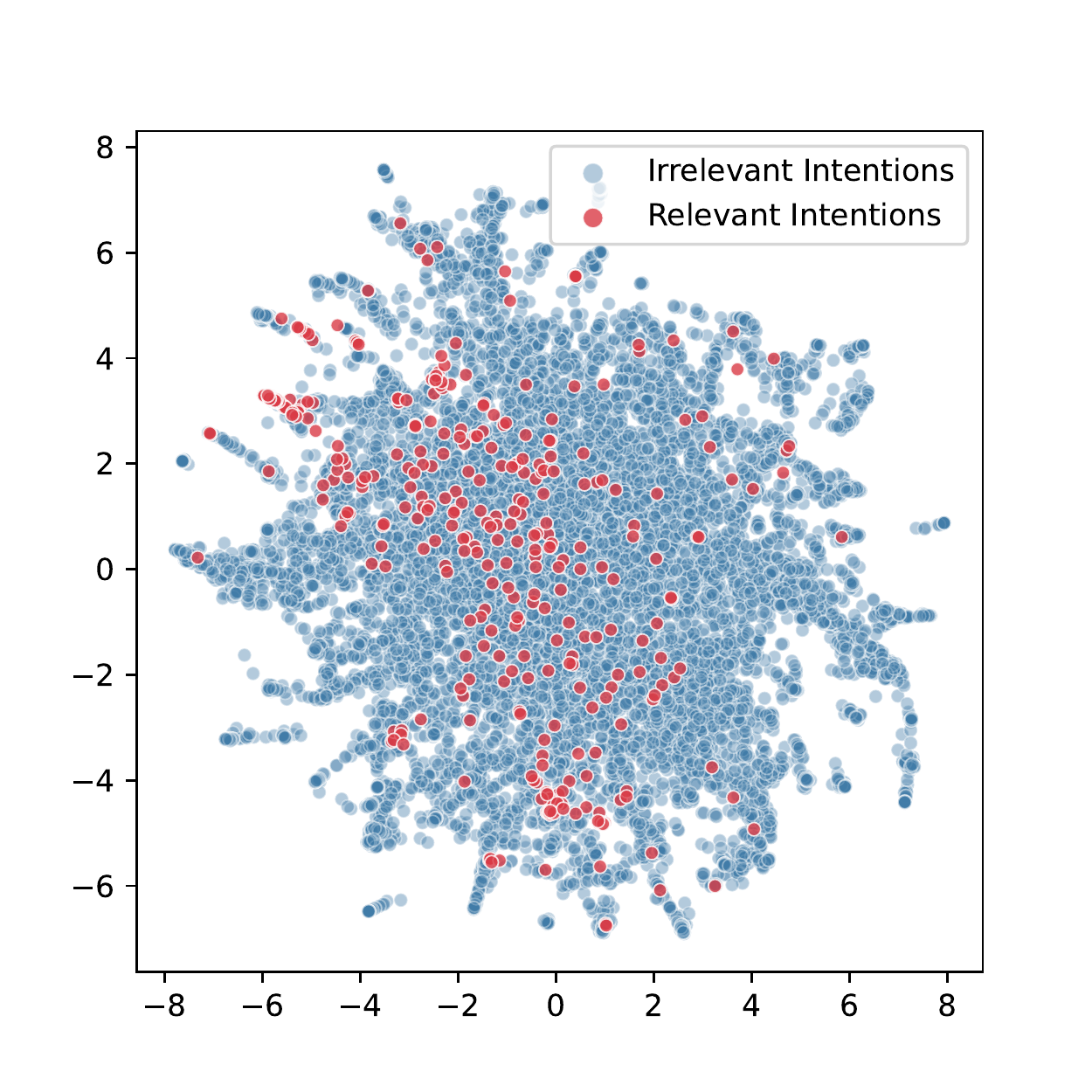}}
\vspace{-2ex}
\caption{T-SNE visualization of mission embeddings trained by SASRec, STOSA, and \ours. \textcolor{blue}{Blue points} are irrelevant points to a certain intention. \textcolor{red}{Red points} are relevant shopping intentions.}\label{fig:tsne}
\vspace{-2ex}
\end{figure*}

\begin{table}[tb]
\caption{Ablation studies on stochastic embeddings on original sequences.}\label{table:ablation-embed}
\centering
\fontsize{7.5}{9.5}\selectfont \setlength{\tabcolsep}{0.3em}
\begin{tabular}{@{}lccccc@{}}
\toprule
Methods            & Recall@1 & NDCG@10 & Recall@2 & Recall@5 & Recall@10 \\ \midrule
\ours w/o covar & 0.5516	&	0.4418	&	0.6280	&	0.7030	&	0.7571   \\
\ours w/o G-mean & 0.5685 &	0.4744	&	0.6626	&	0.7739	&	0.8512 \\
\ours w/o G-covar & 0.6704	&	0.5163	&	0.7553	&	0.8359	&	0.8962 \\
\textbf{\ours} & \textbf{0.7061} &	\textbf{0.5297}       &   	\textbf{0.7890}	&	\textbf{0.8623}	&	\textbf{0.9050}    \\\bottomrule
\end{tabular}
\vspace{-3ex}
\end{table}

\begin{table}[t]
\caption{Study on different graph neural networks (GNN) in intention graph regularizer on original sequences.}\label{table:para-gcn}
\centering
\footnotesize
\begin{tabular}{@{}lccccc@{}}
\toprule
Methods            & Recall@1 & NDCG@10 & Recall@2 & Recall@5 & Recall@10 \\ \midrule
Ours (1-layer GCN) & 0.7061   & 0.5297  & 0.7890   & 0.8623   & 0.9050    \\
Ours (2-layer GCN) & 0.4970   & 0.4633  & 0.6302   & 0.7756   & 0.8564    \\ \midrule
Ours (1-layer GAT) & 0.3390   & 0.4660  & 0.5698   & 0.7949   & 0.9202    \\\bottomrule
\end{tabular}
\vspace{-2ex}
\end{table}

\subsection{Ablation Study (RQ3)}

\noindent \textbf{Ablation on Model Structure.} Figure~\ref{fig:abla} presents the ablation study to verify the effects of different components in \ours.
We compare \ours with model variants that remove one of the key components:
(1) \textbf{Ours w/o Graph Regularizer (GR):} We remove the intention graph regularizer and directly send the stochastic representations generated by the stochastic embedding layer into Transformers;
(2) \textbf{Ours w/o Stochastic Representation (SR):} We degrade the stochastic representations into deterministic embeddings for each intention. 
Then, we have the following observations:
(1) Compared with the backbone model, SASRec, stochastic representations can bring significant improvements of $16.48\%$ in Hit@1 and $3.66\%$ in NDCG@10, and graph regularizer can bring $41.85\%$ of Hit@1 and $14.92\%$ of NDCG@10 increase.
Removing either the graph regularizer or the stochastic representation will cause performance drop in all metrics on all three datasets, which shows the efficacy of the components in our model design;
(2) Comparing the performance drops caused by removing different components, we notice that removing the intention relation graph regularizer is more harmful than the stochastic representation. 
This is because some collaborative transitivity relations captured by stochastic representations may be already included in the global intention relation graph.

\noindent \textbf{Ablation on stochastic Representations.}
Table~\ref{table:ablation-embed} shows the ablation studies on stochastic embeddings in \ours.
Our objective is to study the effect of various components of the stochastic embeddings and assess if the use of graph regularization enhances their performance. To achieve this, we conduct evaluations under three distinct experimental settings:
(1) exclusion of graph regularization on the covariance embedding (w/o G-covar); 
(2) exclusion of graph regularization on the mean embedding (w/o G-mean); 
and (3) exclusion of the entire covariance embedding (w/o covar).
The results of our evaluation indicate that the removal of graph regularization on either component of the stochastic embedding results in a significant decrease in performance. 
This is due to the fact that regularizing only one component of the embedding allows for more flexibility in the other component, thereby weakening the overall regularization power. 
Furthermore, the removal of the entire covariance embedding results in a further decline in performance, as the model degenerates into a lower-dimensional deterministic sequential recommendation model.


\subsection{Parameter Studies (RQ4)}

We have performed extensive parameter studies to evaluate the effects of different parameters on our model:

\noindent \textbf{Effect of Hidden Dimensionalities.} 
We study the influence of hidden dimensionalities of mean and covariance embedding, $d$, towards our methods. 
Among the searching range of $[16$, $32$, $64$, $128$, $256$, $512]$, $d=64$ works the best. 
If $d$ is too small, the model cannot encode the shopping intentions well and cannot learn some high-dimensional relation between intentions. 
If $d$ is too large, it will become a problem for model to learn the high-dimensional representations, also causing a performance drop.

\noindent \textbf{Effect of GNN Types.}
Table~\ref{table:para-gcn} presents the results of using various GNNs in the intention relation graph regularizer. The results indicate that the graph convolution network (GCN) outperforms the graph attention network (GAT).
This is because GAT requires additional edge weight learning for dense, noisy graphs. However, our intention relation graph is not dense, and the edge quality is high, rendering the training of additional attention weights unnecessary.
We utilize the MAD metric~\cite{chen2020measuring} to evaluate the smoothness of node representations. The results show that as the number of GCN layers increases from one to two, the MAD drops significantly from 0.774 to 0.379, indicating that the 2-layer GCN produces smoother representations. On the other hand, the performance drop as shown in Table~\ref{table:para-gcn} suggests that the 2-layer GCN is over-smoothing.

\noindent \textbf{Effect of Number of GCN Layers.}
Besides, we also investigate the effect of the number of GCN layers on \ours.
With a single layer of graph convolution, GCN can only gather data from its immediate neighbors.
Information from larger neighborhoods can only be incorporated when numerous GCN layers are applied.
The results in Table~\ref{table:para-gcn} indicate that 1-layer GCNs perform better than multi-layer GCNs.
The reason is that the directly related intentions on the graph are more crucial when identifying the main shopping intentions from consumer historical data, and introducing more distant neighbours on the graph can introduce more noise.



\subsection{Intention Embeddings Comparison (RQ5)}


To better explain the efficacy of the proposed intention relation graph regularizer, we compare the shopping intention representations trained by SASRec, STOSA, and \ours via t-SNE visualization~\cite{van2008visualizing} in Figure~\ref{fig:tsne}.
The red points in the figure represent the relevant shopping intentions to the intention "miniature", while the blue points indicate the irrelevant ones.
From Figure~\ref{fig:tsne-sasrec}, we observe that although SASRec is trained on user historical data to learn some correlations between intentions, the embeddings of relevant shopping intentions are still quite scattered across the latent space.
From Figure~\ref{fig:tsne-stosa}, the red points start to cluster with each other, because the state-of-the-art distribution-based model, STOSA, can capture the collaborative transitivity between intentions, which are ignored by SASRec.
In contrary, from Figure~\ref{fig:tsne-ours}, it is obvious that the related intention embeddings further cluster with each other. This proves that the intention relation graph truly regularizes the stochastic representations, constricting relevant intentions embeds closer to improve \ours performances. 

\section{Conclusion}
We presented \ours, a graph-regularized stochastic Transformer-based model for main shopping intention identification.
\ours first models the shopping intentions as Gaussian distributions and then creates an intention relation graph as prior knowledge to regularize these distributions. 
The regularized stochastic representations will be fed to the Transformer architecture for user main shopping intention identification.
We perform experiments on three different datasets, representing three different scenarios in real life applications.
Extensive experimental results demonstrate \ours significantly improve the main shopping intention identification from users' historical interaction sequences.
In the future, we will change the Transformer architecture to accommodate distribution-based models more effectively.


\bibliographystyle{ACM-Reference-Format}
\balance
\bibliography{mybib}

\appendix
\newpage
\end{document}